\documentclass[english,prl, reprint,longbibliography]{revtex4-1}
\usepackage[T1]{fontenc}
\usepackage[latin9]{inputenc}
\usepackage{color}
\usepackage{babel}
\usepackage{verbatim}
\usepackage{amsmath}
\usepackage{graphicx}
\usepackage{xargs}[2008/03/08]
\usepackage[unicode=true,pdfusetitle,
 bookmarks=true,bookmarksnumbered=false,bookmarksopen=false,
 breaklinks=false,pdfborder={0 0 0},pdfborderstyle={},backref=false,colorlinks=true]
 {hyperref}
\begin{document}

\title{Abrupt transitions between Markovian and non-Markovian dynamics in
open quantum systems}

\author{Shengshi Pang$^{1,2}$}

\author{Todd A. Brun$^{3}$}

\author{Andrew N. Jordan$^{1,2,4}$}

\affiliation{$^{1}$Department of Physics and Astronomy, University of Rochester,
Rochester, New York 14627, USA}

\affiliation{$^{2}$Center for Coherence and Quantum Optics, University of Rochester,
Rochester, New York 14627, USA}

\affiliation{$^{3}$Department of Electrical Engineering, University of Southern
California, California 90089, USA}

\affiliation{$^{4}$Institute for Quantum Studies, Chapman University, Orange,
CA 92866, USA}

\date{\today}
\begin{abstract}
A rapid restoration of the bath state is usually required to induce
Markovian dynamics for an open quantum system, which typically can
be realized only in limits such as weak system-bath coupling and infinitely
large bath. In this work, we investigate the Markovianity of a qubit
system coupled to a single-qubit bath with the qubit bath being continuously
refreshed by quantum cooling. A surprising result is that there exists
a finite threshold for the cooling rate at which the system transitions
from non-Markovian dynamics to Markovian dynamics, which is in sharp
contrast to the usual understanding that Markovian dynamics is an
asymptotic behavior under the Born-Markov approximation. We also study
the time correlation of the bath, and find that the decay rate of
bath time correlation is of the same order as the system evolution
speed. This suggests that quantum Markovian dynamics can exist beyond
the usual short bath correlation limit. %
\end{abstract}
\maketitle
\global\long\def\tr{{\rm Tr}}
\global\long\def\trd#1{\tr|#1|}
\global\long\def\e{{\rm e}}
\global\long\def\i{{\rm i}}
\global\long\def\d{{\rm d}}
\global\long\def\s{{\rm S}}
\global\long\def\b{{\rm B}}
\newcommandx\sx[1][usedefault, addprefix=\global, 1=]{\sigma_{x}^{#1}}
\newcommandx\sy[1][usedefault, addprefix=\global, 1=]{\sigma_{y}^{#1}}
\newcommandx\sz[1][usedefault, addprefix=\global, 1=]{\sigma_{z}^{#1}}
\global\long\def\xs{\sx[\s]}
\global\long\def\ys{\sy[\s]}
\global\long\def\zs{\sz[\s]}
\global\long\def\xb{\sx[\b]}
\global\long\def\yb{\sy[\b]}
\global\long\def\zb{\sz[\b]}
\newcommandx\rs[1][usedefault, addprefix=\global, 1=t]{\rho_{#1}^{\s}}
\newcommandx\rsp[1][usedefault, addprefix=\global, 1=t]{\rho_{#1}^{\prime\s}}
\global\long\def\htot{H_{{\rm tot}}}
\global\long\def\hs{H_{\s}}
\global\long\def\hb{H_{\b}}
\newcommandx\rsb[1][usedefault, addprefix=\global, 1=t]{\rho_{#1}^{\s\b}}
\global\long\def\hi{H_{{\rm int}}}
\newcommandx\hit[1][usedefault, addprefix=\global, 1=t]{\widetilde{H}_{{\rm int}}(#1)}
\newcommandx\ket[1][usedefault, addprefix=\global, 1=0]{|#1\rangle}
\newcommandx\br[1][usedefault, addprefix=\global, 1=0]{\langle#1|}
\newcommandx\kb[1][usedefault, addprefix=\global, 1=0]{|#1\rangle\langle#1|}
\newcommandx\kbt[2][usedefault, addprefix=\global, 1=0, 2=1]{|#1\rangle\langle#2|}
\global\long\def\sm{\sigma_{-}^{\b}}
\global\long\def\sp{\sigma_{+}^{\b}}
\newcommandx\lb[1][usedefault, addprefix=\global, 1=\sm]{\mathcal{D}[#1]}
\newcommandx\rr[1][usedefault, addprefix=\global, 1=\rsb]{\mathcal{R}^{\s}[#1]}
\newcommandx\pt[1][usedefault, addprefix=\global, 1=t]{\partial_{#1}}
\global\long\def\delt{\Delta t}
\global\long\def\xo{x_{0}}
\global\long\def\yo{y_{0}}
\global\long\def\zo{z_{0}}
\newcommandx\rb[1][usedefault, addprefix=\global, 1=t]{\rho_{#1}^{\b}}
\newcommandx\apt[1][usedefault, addprefix=\global, 1=t]{\alpha_{#1}}
\newcommandx\bet[1][usedefault, addprefix=\global, 1=t]{\beta_{#1}}
\global\long\def\lp{\lambda^{\prime}}
\global\long\def\ib{I^{\b}}
\newcommandx\aij[1][usedefault, addprefix=\global, 1=ij]{a_{#1}(t)}
\newcommandx\si[1][usedefault, addprefix=\global, 1=\s]{\sigma_{i}^{#1}}
\newcommandx\sj[1][usedefault, addprefix=\global, 1=\b]{\sigma_{j}^{#1}}
\newcommandx\aijd[1][usedefault, addprefix=\global, 1=ij]{\dot{a}_{#1}(t)}
\global\long\def\ob{0^{\b}}
\global\long\def\is{I^{\s}}
\global\long\def\os{0^{\s}}
\global\long\def\zop{z_{0}^{\prime}}
\global\long\def\xop{x_{0}^{\prime}}
\global\long\def\yop{y_{0}^{\prime}}
\newcommandx\del[1][usedefault, addprefix=\global, 1=]{\Delta_{#1}}
\global\long\def\dels#1{\Delta_{#1}^{2}}
\newcommandx\dtr[2][usedefault, addprefix=\global, 1=, 2=]{d_{\tr}(#1,#2)}
\global\long\def\fbt{F_{t}^{\b}}
\global\long\def\avg#1{\langle#1\rangle_{\b}}
\global\long\def\xl{x_{1}}
\global\long\def\yl{y_{1}}
\global\long\def\zl{z_{1}}
\newcommandx\baco[2][usedefault, addprefix=\global, 1=t+\tau, 2=t]{\langle\widetilde{\xb}(#1)\widetilde{\xb}(#2)\rangle}
\newcommandx\li[1][usedefault, addprefix=\global, 1=i]{\lambda_{#1}}
\global\long\def\sysy{\sy\otimes\sy}
\newcommandx\sgn[1][usedefault, addprefix=\global, 1=\xi]{{\rm sgn}(#1)}
\newcommandx\ct[1][usedefault, addprefix=\global, 1=t]{c_{#1}}
\newcommandx\dct[1][usedefault, addprefix=\global, 1=t]{\dot{c}_{#1}}
\newcommandx\lt[2][usedefault, addprefix=\global, 1=0, 2=t]{\Lambda_{#1\rightarrow#2}}
\newcommandx\rst[2][usedefault, addprefix=\global, 1=, 2=t]{\widetilde{\rho}_{#2}^{(#1)}}
\newcommandx\cts[1][usedefault, addprefix=\global, 1=t]{c_{#1}^{2}}
\newcommandx\tk[1][usedefault, addprefix=\global, 1=]{t_{#1}}
\newcommandx\rk[2][usedefault, addprefix=\global, 1=, 2=t]{\rho_{#1}(#2)}
\global\long\def\mn{\mathcal{N}}
\global\long\def\tp{t^{\prime}}
\global\long\def\dtp{\d\tp}
\newcommandx\xt[1][usedefault, addprefix=\global, 1=t]{x_{#1}}
\newcommandx\yt[1][usedefault, addprefix=\global, 1=t]{y_{#1}}
\newcommandx\zt[1][usedefault, addprefix=\global, 1=t]{z_{#1}}
\global\long\def\uo{u_{0}}
\newcommandx\rt[1][usedefault, addprefix=\global, 1=t]{\rho_{#1}}
\global\long\def\trb{\tr_{\b}}
\newcommandx\ba[2][usedefault, addprefix=\global, 1=\alpha, 2=t]{B_{#1}(#2)}
\newcommandx\bad[2][usedefault, addprefix=\global, 1=\alpha, 2=t]{B_{#1}^{\dagger}(#2)}
\newcommandx\ft[1][usedefault, addprefix=\global, 1={t,\tp}]{f_{\b}(#1)}
\newcommandx\vsb[1][usedefault, addprefix=\global, 1=t]{\boldsymbol{v}_{#1}^{\s\b}}
\newcommandx\vs[1][usedefault, addprefix=\global, 1=t]{\boldsymbol{v}_{#1}^{\s}}
\global\long\def\wo{w_{0}}
\newcommandx\qt[2][usedefault, addprefix=\global, 1=t, 2=]{Q_{#1}^{#2}}
\newcommandx\dqt[1][usedefault, addprefix=\global, 1=t]{\dot{Q}_{#1}}
\newcommandx\sij[1][usedefault, addprefix=\global, 1=ij]{s_{#1}}
\newcommandx\rsbt[1][usedefault, addprefix=\global, 1=t]{\widetilde{\rho}_{#1}^{\s\b}}
\newcommandx\rbt[1][usedefault, addprefix=\global, 1=t]{\widetilde{\rho}_{#1}^{\b}}
\renewcommandx\ft[1][usedefault, addprefix=\global, 1={t,\tp}]{f_{\b}(#1)}
\global\long\def\dn{\Delta_{n}}
\newcommandx\tn[1][usedefault, addprefix=\global, 1=n]{t_{#1}}
\newcommandx\dx[1][usedefault, addprefix=\global, 1=]{\Delta_{x}^{#1}}
\newcommandx\dy[1][usedefault, addprefix=\global, 1=]{\Delta_{y}^{#1}}
\newcommandx\dz[1][usedefault, addprefix=\global, 1=]{\Delta_{z}^{#1}}
\global\long\def\vr{\boldsymbol{r}_{0}}
\global\long\def\vrp{\boldsymbol{r}_{0}^{\prime}}

\section{Introduction}

The dynamics of an open quantum system is very different from the
unitary evolution of a closed quantum system, as it may lose information
from the system to the environment and become physically irreversible.
 Such dynamics cannot be simply described by the ordinary Schr\"odinger
equation, and needs more sophisticated tools such as the Feynman-Vernon
influence functional \cite{Feynman1963}. The interest in open quantum
systems has grown more intensive in recent years with the development
of quantum information, as it is important to learn how quantum technologies
can work in real environments.

Regardless of the physical details, the dynamics of open quantum systems
can be roughly divided into two categories based on the memory effect
of the bath: Markovian and non-Markovian dynamics. If the bath correlation
time is much shorter than the time scale of the system evolution,
the bath has almost no memory effect and the state of the system at
any time is determined by its state at the immediate previous time
step, and the evolution of the system can be effectively described
by a dynamical semigroup \cite{Davies1974,Lindblad1976,Davies1976}.
This kind of open system dynamics is called Markovian. By contrast,
if the bath correlation is enduring and lasts for a time comparable
to the time scale of the system evolution, the instantaneous state
of the system is determined by the entire history of the system evolution
rather than its immediate predecessor, and the dynamics of the system
do not form a semigroup generally. This kind of open system dynamics
is called non-Markovian.

While Markovian quantum dynamics has been studied with various approximations
for a long time, the exact quantification or even definition of Markovianity
for quantum dynamics is not easy. Two main approaches to defining
the Markovianity of quantum dynamics currently are completely positive
(CP) divisibility \cite{Wolf2008,Rivas2010,Rivas2011} and distinguishability
of quantum states \cite{Breuer2009,Breuer2016}. The CP-divisibility
approach defines Markovian quantum dynamics as those which can be
decomposed into a sequence of completely positive maps, representing
the evolution for successive, arbitrarily-chosen intervals of time.
The distinguishability of quantum states approach defines Markovian
quantum dynamics as those under which the trace distance between two
arbitrary quantum states always decreases with time, which can be
interpreted as irreversible loss of information from the system into
the environment. A notable recent advance is that these two approaches
were proven to be equivalent, first for quantum states in special
Hilbert spaces \cite{Bylicka2017} and then universally \cite{Buscemi2016,Chruscinski2017a}.
Meanwhile, many different measures have been proposed to witness or
quantify the Markovianity the dynamics of an open quantum system,
among which are quantum Fisher information flow \cite{Lu2010}, quantum
fidelity \cite{Rajagopal2010}, quantum correlation flow \cite{Luo2012},
quantum channel capacity \cite{Bylicka2014}, geometry of dynamically
accessible states \cite{Lorenzo2013}, quantum interferometric power
\cite{Dhar2015}, etc; and the Lindblad master equation has been modified
in various ways to include memory effects \cite{Daffer2004,Shabani2005}.
We refer the readers to \cite{Rivas2014,Koch2016,Breuer2016,deVega2017}
for reviews of recent progress in open quantum systems.

A common and widely held assumption in Markovian quantum dynamics
is that fast restoration of the bath state is required, so that the
evolution of the quantum system is memoryless, which is rooted in
the Born-Markov approximation. If the bath state is not restored fast
enough, the information of the system lost into the bath may flow
back into the system, and the dynamics of the system becomes non-Markovian.

An interesting question is: how rapidly does the bath state need to
be restored to make the dynamics of an open quantum system Markovian?

In this paper, we study this problem by a simple model with a qubit
as the system and another qubit as the bath. The bath qubit is assumed
to be in the ground state initially, and is continuously cooled down
to that ground state to simulate the restoration process. In principle,
the cooling process requires another large reservoir at zero temperature,
and the interaction between the large reservoir and the bath qubit
here can be complex. However, as we do not care about the details
of the cooling process, we can trace out the degrees of freedom of
the large reservoir and effectively formulate the cooling of the bath
qubit as a simple energy decay process \cite{Breuer2007a,Gardiner2010}.

As the dimension of bath is low in this problem, one may expect a
strong memory effect from the bath on the system and an extremely
fast restoration of the bath state needed to make the system dynamics
Markovian. However, the result reveals that the bath does not need
to be refreshed extremely fast to make the system dynamics Markovian,
and interestingly there exists a sharp threshold in the bath cooling
rate between the non-Markovian regime and the Markovian regime of
the system dynamics, and the threshold is of the same order of the
system evolution speed. Moreover, calculating the bath correlation
function shows that the decay rate of bath correlation is just the
bath cooling rate, implying that the dynamics of the system can become
Markovian even when the bath correlation time is comparable to the
time scale of system evolution. This is in marked contrast to the
usual belief that the Markovian dynamics is an asymptotic behavior
in the limit of short bath correlation times.

\section{Markovian quantum dynamics}

In an open quantum system, the system interacts with a bath. The total
Hamiltonian of the system and the bath is
\begin{equation}
\htot=\hs+\hb+\hi,
\end{equation}
where $\hs$, $\hb$ are the free Hamiltonians of the system and the
bath respectively and $\hi$ is the interaction Hamiltonian between
the system and the bath.

To derive the reduced dynamics of the system, one can transform into
the interaction picture, in which the interaction Hamiltonian becomes
$\hit=\e^{\i t(\hs+\hb)}\hi\e^{-\i t(\hs+\hb)}$, and make the Born-Markov
approximation. The Born approximation assumes that the joint state
of the system and the bath at any time $t$ can be written as $\rsb\approx\rs\otimes\rb[0]$,
assuming that the coupling between the system and the bath is weak
and the size of the bath is large, where $\rs$ is the reduced density
matrix of the system at time $t$ and $\rb[0]$ is the reduced density
matrix of the bath at the initial time. One can then obtain the master
equation for the system \cite{Breuer2007a}:
\begin{equation}
\pt\rs=-\int_{0}^{\infty}\trb[\hit,[\hit[t-\tp],\rs[\tp]\otimes\rb[0]]]\dtp,\label{eq:gen mas eq}
\end{equation}
which assumes $\trb[\hit,\rb[0]]=0$. If the interaction Hamiltonian
$\hit$ can be decomposed as $\hit=\sum_{\alpha}A_{\alpha}(t)\otimes\ba$,
the partial trace of the bath degrees of freedom in Eq. (\ref{eq:gen mas eq})
will generate the bath correlation functions
\begin{equation}
\avg{{{\bad}}{{\ba[\beta][t-\tp]}}}=\trb\Big[\bad\ba[\beta][t-\tp]\rb[0]\Big].\label{eq: bath corr}
\end{equation}
Eq. (\ref{eq:gen mas eq}) shows that the evolution of $\rs$ generally
depends on the history of $\rs$. To remove the time integral in (\ref{eq:gen mas eq})
and make the equation Markovian, the bath correlation functions (\ref{eq: bath corr})
need to decay very fast with time so that the contributions from the
state at times far from the current time $t$ in the integral are
negligible and the time variation of the system state depends only
on the system state at time $t$. In this case, the $\rs[\tp]$ in
the integral in Eq. (\ref{eq:gen mas eq}) can be replaced by $\rs$,
which is called the Markov approximation. The dynamics of the system
described by such a Markovian master equation is generally coarse-grained
in the time framework and cannot be resolved within the bath correlation
time, but the faster the bath correlation functions decay, the more
precisely the system dynamics can be described by the master equation.

\emph{Model.}\textemdash{} In this work, we are interested in how
fast the bath state needs to be refreshed in order to make the system
dynamics Markovian. This is important to the understanding of quantum
Markovian dynamics, as a fast restoration of the bath state is one
of the key assumptions in the Born-Markov approximation reviewed above.
We consider a simple model with a qubit as the system and another
qubit as the bath. We suppose the system and the bath interact via
the $X$-$X$ coupling, 
\begin{equation}
\hi=\xi\xs\otimes\xb.
\end{equation}
For simplicity, we drop the free Hamiltonian of both the system and
the bath.

The dynamics of the system qubit is generally non-Markovian as it
is coupled to a bath. To rapidly restore the bath state, we prepare
the bath in the ground state initially and introduce a continuous
cooling of the bath to that ground state. The cooling process can
be described by a dissipation term $\kappa\lb\rsb\delt$, where $\kappa$
is the cooling rate and $\rsb$ is the density matrix of the joint
state of the system and bath at time $t$. Putting the Hamiltonian
evolution and the cooling process together, we have the following
master equation for the joint state of the system qubit and the bath
qubit:
\begin{equation}
\pt\rsb=-\i\xi[\xs\otimes\xb,\rsb]+\kappa\lb\rsb,\label{eq:master eq}
\end{equation}
where $\lb$ is performed on the bath qubit alone, defined as $\lb\rsb=\sm\rsb\sp-\frac{1}{2}\{\sp\sm,\,\rsb\}$.
In principle, there should be another term $\lb[\sp]\rsb$ in Eq.
(\ref{eq:master eq}) representing the heating process, if the thermal
reservoir in contact with the qubit bath is not at zero temperature
\cite{Gardiner2010}, due to the detailed balance principle. Here,
as we want to cool the bath to the ground state, we assume the thermal
reservoir to be at the zero temperature, and the heating term vanishes
in this case.

To study the Markovianity of the system dynamics, we need to find
how the system evolves. The derivation of the evolution of an open
quantum system is generally non-trivial. However, in this problem,
as the system and the bath are simple enough, we can obtain the exact
evolution of the system by solving the master equation (\ref{eq:master eq})
exactly.

Let us assume the initial state of the system qubit is $\rs[0]=\frac{1}{2}\big(\is+\xo\xs+\yo\ys+\zo\zs\big)$
and the initial state of the bath is $\ket[\ob]$ which is also the
state that the bath will be cooled down to. It is straightforward
to obtain the solution to the joint state of the system and the bath
at any time $t$ from Eq. (\ref{eq:master eq}). If we trace out the
degrees of freedom of the bath and focus on the state of the system
alone, we can obtain the reduced density matrix of the system qubit
at any time $t$, which turns out to be
\begin{equation}
\rs=\frac{1}{2}\Big[\is+\xo\xs+\ct\big(\yo\ys+\zo\zs\big)\Big],\label{eq:sys map}
\end{equation}
where $\ct=\e^{-\frac{\kappa t}{4}}\Big(\frac{\kappa\sinh\frac{t}{4}\sqrt{\kappa^{2}-64\xi^{2}}}{\sqrt{\kappa^{2}-64\xi^{2}}}+\cosh\frac{t}{4}\sqrt{\kappa^{2}-64\xi^{2}}\Big)$.
This implicitly assumes that $\kappa^{2}>64\xi^{2}$. If $\kappa^{2}<64\xi^{2}$,
$\ct$ should be $\e^{-\frac{\kappa t}{4}}\Big(\frac{\kappa\sin\frac{t}{4}\sqrt{64\xi^{2}-\kappa^{2}}}{\sqrt{64\xi^{2}-\kappa^{2}}}+\cos\frac{t}{4}\sqrt{64\xi^{2}-\kappa^{2}}\Big)$;
and if $\kappa^{2}=64\xi^{2}$, $\ct=\e^{-\frac{\kappa t}{4}}\big(1+\frac{1}{4}\kappa t\big)$.

This is the solution to the system dynamics. It is simple, but will
turn out to give an interesting characterization of the Markovianity
of the system dynamics.

\section{Completely positive divisibility}

Based on the evolution of the system state above, we can study the
Markovianity of the evolution of the system. We will first look at
the CP-divisibility of the dynamics, and later consider quantum state
distinguishability.

The CP-divisibility criterion \cite{Rivas2010}\emph{ }tells that
a quantum dynamics $\lt$ is Markovian if each piece $\lt[{\tk[i]}][{\tk[i+1]}]$
of an arbitrary division of $\lt$,
\begin{equation}
\lt[][t]=\lt[{\tk[n]}][t]\lt[{\tk[n-1]}][{\tk[n]}]\cdots\lt[0][{\tk[1]}],
\end{equation}
is completely positive. An intuitive idea behind this definition is
that if the evolution of a quantum system between two arbitrary times
is completely positive, and hence can be derived from an interaction
with a memoryless bath, then the entire evolution is memoryless and
depends only on the state of the system at the present time.

Determining the CP-divisibility of a quantum dynamics is highly non-trivial
in general. However, it can be proven \cite{Rivas2011} that a quantum
dynamics is CP-divisible if and only if the system evolution can be
formulated as a master equation with the coefficients of the dissipation
terms non-negative for all times. This gives a more convenient way
to determine whether a quantum dynamics is CP-divisible or not. Below,
we will use this theorem to figure out when the evolution of the system
qubit in the current problem is CP-divisible.

We first need to find the master equation for the evolution of the
system qubit. In the Appendix, it was obtained that the master equation
for the system evolution corresponding to (\ref{eq:sys map}) is
\begin{equation}
\pt\rs=-\frac{\dct}{2\ct}\lb[\xs]\rs.\label{eq:qubit master eq}
\end{equation}

From the solution of $\ct$ in the last section, one can see that
if $\kappa^{2}>64\xi^{2}$, $-\frac{\dct}{2\ct}=\frac{8\xi^{2}}{\kappa+\sqrt{\kappa^{2}-64\xi^{2}}\coth\big(\frac{1}{4}t\sqrt{\kappa^{2}-64\xi^{2}}\big)}$,
which is always positive; if $\kappa^{2}<64\xi^{2}$, $-\frac{\dct}{2\ct}=\frac{8\xi^{2}}{\kappa+\sqrt{64\xi^{2}-\kappa^{2}}\cot\big(\frac{1}{4}t\sqrt{64\xi^{2}-\kappa^{2}}\big)},$which
is negative when $\cot\big(\frac{1}{4}t\sqrt{64\xi^{2}-\kappa^{2}}\big)<-\frac{\kappa}{\sqrt{64\xi^{2}-\kappa^{2}}}$.
And if $\kappa^{2}=64\xi^{2}$, $-\frac{\dct}{2\ct}=\frac{\kappa^{2}t}{2(16+4\kappa t)},$
which is non-negative for all times $t$.

\begin{figure}
\includegraphics[scale=0.6]{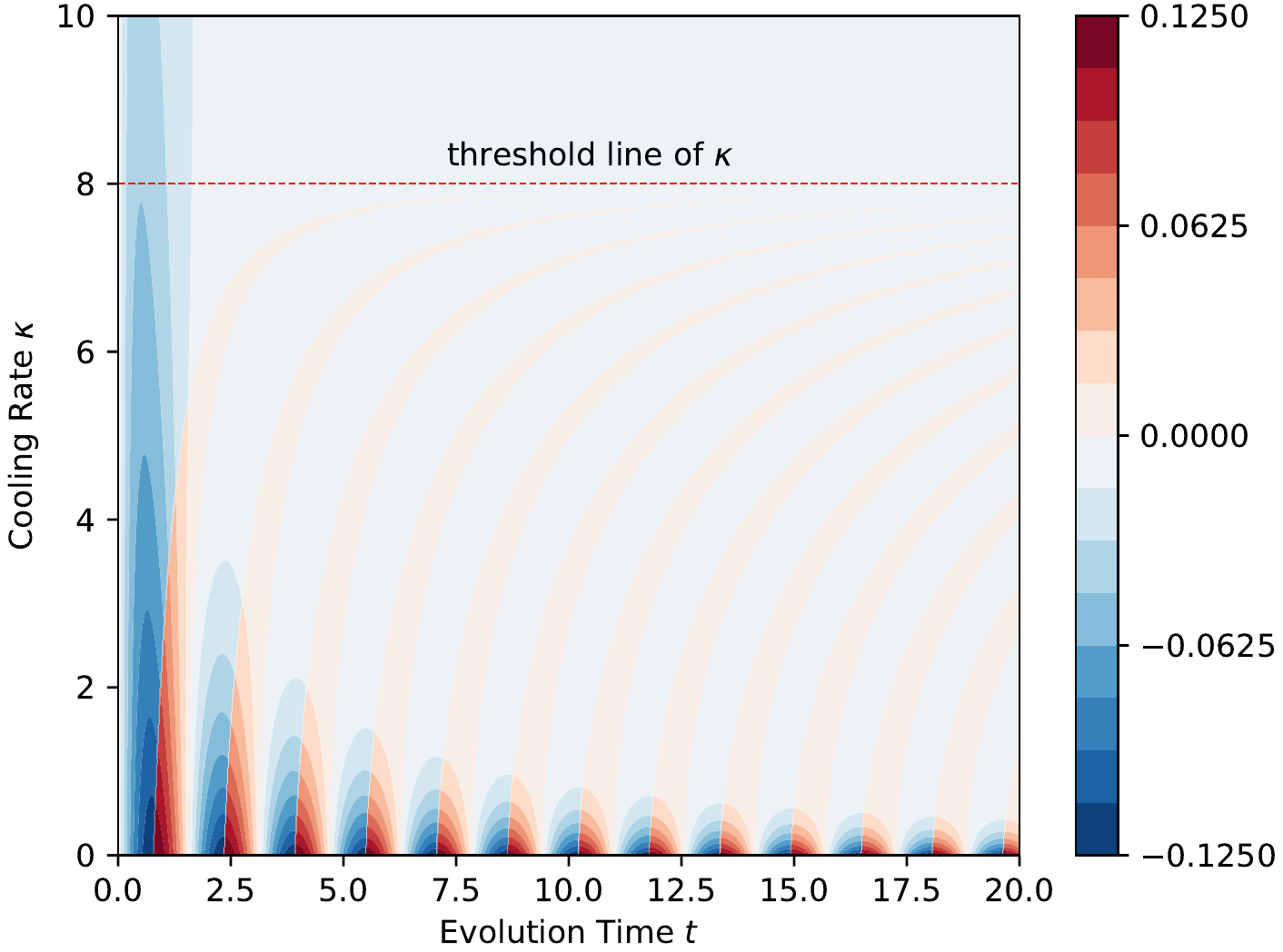}\caption{\label{fig:Contour}(Color online) Contour plot of $\protect\pt|\protect\ct|$.
The interaction strength $\xi$ between the system qubit and the bath
qubit is set to $1$. It can be seen that when $\kappa\geq8$, $\protect\pt|\protect\ct|$
is always non-positive, while when $\kappa<8$, $\protect\pt|\protect\ct|$
can be positive periodically although the time intervals of positive
$\protect\pt|\protect\ct|$ decrease with $\kappa$. As $\protect\sgn[{{\protect\dct}}/{{\protect\ct}}]=\protect\sgn[{{\protect\pt}}|{{\protect\ct}}|]$,
the figure also implies that $\frac{\protect\dct}{\protect\dct}$
is always non-positive when $\kappa\geq8$ while $\frac{\protect\dct}{\protect\ct}$
has oscillation in its sign when $\kappa<8$. It means that the dynamics
of the system qubit is Markovian when $\kappa\geq8$ and becomes non-Markovian
when $\kappa<8$, in both the CP-divisibility and the information
theoretic criteria. $\kappa=8$ is thus the threshold of bath cooling
rate that divides the Markovian and non-Markovian regimes of the system
qubit dynamics.}
\end{figure}

Therefore, we can immediately infer by the CP-divisibility theorem
that the dynamics of the system qubit is Markovian when $\kappa^{2}\geq64\xi^{2}$,
and is non-Markovian when $\kappa^{2}<64\xi^{2}$. We plot $\pt|\ct|$
for different cases in Fig. \ref{fig:Contour}. $\pt|\ct|$ has the
same sign as $\frac{\dct}{\ct}$, so it shows how the sign of $\frac{\dct}{\ct}$
varies with time for different cooling rates, which verifies the above
results.

The interesting thing here is that there exists a threshold line $\kappa^{2}=64\xi^{2}$
that divides the Markovian and non-Markovian regimes of the system
dynamics. When $\kappa$ goes from above $8|\xi|$ to below $8|\xi|$,
the system undergoes an abrupt transition from non-Markovian dynamics
to Markovian dynamics. More importantly, this transition line lies
at the same order of the $\xi$, which means that the system dynamics
becomes Markovian when the bath restoration rate is of the same order
as the system evolution speed. This shows the existence of Markovian
quantum dynamics beyond the traditional asymptotic regime of $\kappa\gg\xi$
under the Born-Markov approximation.

\section{Distinguishability of quantum states}

The non-Markovianity of quantum dynamics can also be defined and measured
in an information theoretic way, through the distinguishability of
two different quantum states. The trace distance between two quantum
states determines the distinguishability of two quantum states \cite{Holevo1973,Helstrom1976},
which can be considered as a measure of the information that can be
learned from the two states. Breuer \emph{et al.} pointed out that
if a quantum dynamics is Markovian, the trace distance between any
two quantum states will decrease with time \cite{Breuer2007a}, due
to the contractivity of completely positive quantum maps. The non-Markovianity
of quantum dynamics can be defined as the existence of a pair of quantum
states whose trace distance increases for some time under that dynamics
\cite{Breuer2009}. An intuitive interpretation for this definition
is that if the trace distance between two quantum states increases
for some time during the evolution, it means that some of the information
about the system that was lost into the environment flows back into
the system. This implies that the environment has memory, and thus
the dynamics of the quantum system is non-Markovian.

The trace distance between two quantum states $\rk[1]$, $\rk[2]$
is defined as
\begin{equation}
\dtr[{\rk[1]}][{\rk[2]}]=\frac{1}{2}\trd{{{\rk[1]}}-{{\rk[2]}}}.
\end{equation}
A relationship between the trace distance and the distinguishability
of $\rk[1]$, $\rk[2]$ is that the maximum probability of correctly
distinguishing $\rk[1]$ and $\rk[2]$ is $\frac{1}{2}\big[1+\dtr[{\rk[1]}][{\rk[2]}]\big]$.
Therefore, the trace distance determines the distinguishability between
two quantum states, and characterizes how much information one may
extract from the two states. 

Now, we can use this information theoretic definition of quantum non-Markovianity
to examine the Markovianity of the quantum dynamics in our problem.
Suppose we have two arbitrary initial states for the system qubit,
$\rs[0]=\frac{1}{2}(\is+\xo\xs+\yo\ys+\zo\zs)$ and $\rsp[0]=\frac{1}{2}(\is+\xop\xs+\yop\ys+\zop\zs)$.
It follows from Eq. (\ref{eq:sys map}) that $\rs-\rsp=\frac{1}{2}(\del[x]\xs+\ct\del[y]\ys+\ct\del[z]\zs)$,
and the trace distance between $\rs$ and $\rsp$ is $\dtr[{\rs}][{\rsp}]=\frac{1}{2}\sqrt{\dels x+\cts\left(\dels y+\dels z\right)}$,
where $\del[x]=\xo-\xop$, $\del[y]=\yo-\yop$, $\del[z]=\zo-\zop$.
To see whether the trace distance between $\rs$ and $\rsp$ increases
or not during the evolution of the system, we can take the derivative
of $\dtr[{\rs}][{\rsp}]$ with respect to time $t$, and it produces
$\pt\dtr[{\rs}][{\rsp}]=\left(\dels y+\dels z\right)\frac{\ct\dct}{\dtr[{\rs}][{\rsp}]}$.

Considering $\sgn[{{\ct}}{{\dct}}]=\sgn[\frac{{{\dct}}}{{{\ct}}}]$,
the time derivative of trace distance gives the same characterization
of the system dynamics as the CP-divisibility: when $\kappa^{2}\geq64\xi^{2}$,
the trace distance always decreases with time and the system dynamics
(\ref{eq:sys map}) is Markovian; when $\kappa^{2}<64\xi^{2}$, the
trace distance with $\del[y]\neq0$ or $\del[z]\neq0$ can increase
at times $t$ that satisfies $\cot\left(\frac{1}{4}t\sqrt{64\xi^{2}-\kappa^{2}}\right)<-\frac{\kappa}{\sqrt{64\xi^{2}-\kappa^{2}}}$
, so the system dynamics is non-Markovian. Fig. \ref{fig:Contour}
illustrates this result clearly.

Moreover, we can take the integral of the increase in trace distance
during the whole evolution of the system as the measure of non-Markovianity,
which was proposed by Breuer \emph{et al.} \cite{Breuer2009},
\begin{equation}
\mn=\max_{\rk[1,2][0]}\int_{\pt[\tp]d_{\tr}>0}\pt[\tp]\dtr[{\rk[1][\tp]}][{\rk[2][\tp]}]\dtp,\label{eq:non-markov meas}
\end{equation}
where the integral is taken over all time intervals during which the
trace distance $d_{\tr}$ increases. This measure of non-Markovianity
is non-zero only for the case $\kappa^{2}<64\xi^{2}$, as there is
increase in the trace distance only in that region. The result turns
out to be (see Appendix)
\begin{equation}
\mn={\displaystyle \frac{1}{\exp\Big(\frac{\kappa\pi}{\sqrt{64\xi^{2}-\kappa^{2}}}\Big)-1}}.\label{eq:N}
\end{equation}

It can be clearly seen from the measure of non-Markovianity (\ref{eq:non-markov meas})
that when there is no cooling on the bath qubit, $\kappa=0$, $\mn$
is infinity, at its maximum, as there is periodic information backflow
from the bath to the system and the amplitude of the backflow never
decays, indicating the strongest non-Markovianity for this case. When
$\kappa$ becomes nonzero, $\mn$ becomes finite, as the amplitude
of the information backflow decays with time, indicating a weaker
non-Markovianity. When $\kappa^{2}=64\xi^{2}$, $\mn=0$, indicating
the vanishing of non-Markovianity.

\section{Bath correlation time}

Finally, we want to investigate the bath correlation for this problem.
The bath correlation time plays a key role in the Born-Markov approximation
for Markovian quantum dynamics, as the main characteristic of Markovian
dynamics is that the bath correlation decays extremely fast so that
the system dynamics is memoryless. And in this case, the Lindblad
master equation can be derived in the weak coupling limit \cite{Breuer2007a}
which is the most widely-used approach to describing Markovian quantum
dynamics. 

Below we study the bath correlation to compare the bath correlation
time with the time scale of the system evolution. In fact, as the
bath qubit is being continuously cooled to the ground state at rate
$\kappa$, one can expect that the bath correlation time is approximately
$1/\kappa$. In the following, we will see this is indeed the case.

We first perform a non-unitary transformation of the bath state to
eliminate the dissipation term in the master equation: $\rsb\rightarrow\e^{-\kappa t\lb}\rsb$.
{} In the new picture under this transformation, we can trace out the
degrees of freedom of the bath and obtain a reduced master equation
for the system qubit similar to (\ref{eq:gen mas eq}). The period
over which earlier states of the system can influence the current
state of the system is determined by the time width of bath correlation
functions.

In the Appendix, the relevant bath correlation function between two
arbitrary time $t$ and $\tp$, $t\geq\tp$, is derived, which turns
out to be
\begin{equation}
\ft=\trb\Big[\xb\e^{\kappa(t-\tp)\lb}(\xb\rb[0]+\rb[0]\xb)\Big].
\end{equation}
In this problem, $\rb[0]=\kb[\ob]$, so $\xb\rb[0]+\rb[0]\xb=\xb$.
It is straightforward to obtain that $\e^{\kappa(t-\tp)\lb}\xb=\e^{-\frac{\kappa}{2}(t-\tp)}\xb$.
Therefore, the bath correlation function is
\begin{equation}
\ft=\exp\Big[-\frac{1}{2}\kappa(t-\tp)\Big].\label{eq:bath corr}
\end{equation}
It is homogeneous with time, only depending on the interval between
two times, and decays with the time interval $t-\tp$ at rate $\frac{\kappa}{2}$,
which verifies our previous intuition (up to the factor $\frac{1}{2}$).

As we have known from the previous results that the system dynamics
becomes Markovian when $\kappa\geq8|\xi|$, the bath correlation function
(\ref{eq:bath corr}) implies that Markovian quantum dynamics can
exist with a bath correlation time comparable to the time scale of
the system evolution.

\section{Conclusions}

In the usual Markovian dynamics under the Born-Markov approximation,
the bath must restore much faster than the system evolves so that
the bath is memoryless and the evolution of the system is determined
only by its current state. In this work, however, a surprising result
is that the system dynamics can be Markovian when the bath restoration
rate is of the same order as the system evolution rate, which implies
the existence of Markovian quantum dynamics beyond the limit of fast
bath restoration or short bath correlation. Moreover, a sharp boundary
exists in the bath restoration rate, over which the system undergoes
an abrupt transition between non-Markovian and Markovian dynamics,
which does not show under the Born-Markov approximation.

\section*{Acknowledgments}

S. Pang and A. N. Jordan acknowledge support from the US Army Research
Office under Grant No. W911NF-15-1-0496 and support from the National
Science Foundation under Grant No. DMR-1506081. T. A. Brun acknowledges
support from the National Science Foundation under Grant No. QIS-1719778.

\bibliographystyle{apsrev4-1}
\bibliography{nonmarkovian}

\begin{thebibliography}{27}%
\makeatletter
\providecommand \@ifxundefined [1]{%
 \@ifx{#1\undefined}
}%
\providecommand \@ifnum [1]{%
 \ifnum #1\expandafter \@firstoftwo
 \else \expandafter \@secondoftwo
 \fi
}%
\providecommand \@ifx [1]{%
 \ifx #1\expandafter \@firstoftwo
 \else \expandafter \@secondoftwo
 \fi
}%
\providecommand \natexlab [1]{#1}%
\providecommand \enquote  [1]{``#1''}%
\providecommand \bibnamefont  [1]{#1}%
\providecommand \bibfnamefont [1]{#1}%
\providecommand \citenamefont [1]{#1}%
\providecommand \href@noop [0]{\@secondoftwo}%
\providecommand \href [0]{\begingroup \@sanitize@url \@href}%
\providecommand \@href[1]{\@@startlink{#1}\@@href}%
\providecommand \@@href[1]{\endgroup#1\@@endlink}%
\providecommand \@sanitize@url [0]{\catcode `\\12\catcode `\$12\catcode
  `\&12\catcode `\#12\catcode `\^12\catcode `\_12\catcode `\%12\relax}%
\providecommand \@@startlink[1]{}%
\providecommand \@@endlink[0]{}%
\providecommand \url  [0]{\begingroup\@sanitize@url \@url }%
\providecommand \@url [1]{\endgroup\@href {#1}{\urlprefix }}%
\providecommand \urlprefix  [0]{URL }%
\providecommand \Eprint [0]{\href }%
\providecommand \doibase [0]{http://dx.doi.org/}%
\providecommand \selectlanguage [0]{\@gobble}%
\providecommand \bibinfo  [0]{\@secondoftwo}%
\providecommand \bibfield  [0]{\@secondoftwo}%
\providecommand \translation [1]{[#1]}%
\providecommand \BibitemOpen [0]{}%
\providecommand \bibitemStop [0]{}%
\providecommand \bibitemNoStop [0]{.\EOS\space}%
\providecommand \EOS [0]{\spacefactor3000\relax}%
\providecommand \BibitemShut  [1]{\csname bibitem#1\endcsname}%
\let\auto@bib@innerbib\@empty
\bibitem [{\citenamefont {Feynman}\ and\ \citenamefont
  {Vernon}(1963)}]{Feynman1963}%
  \BibitemOpen
  \bibfield  {author} {\bibinfo {author} {\bibfnamefont {R.~P.}\ \bibnamefont
  {Feynman}}\ and\ \bibinfo {author} {\bibfnamefont {F.~L.}\ \bibnamefont
  {Vernon}},\ }\href@noop {} {\bibfield  {journal} {\bibinfo  {journal} {Ann.
  Phys.}\ }\textbf {\bibinfo {volume} {24}},\ \bibinfo {pages} {118} (\bibinfo
  {year} {1963})}\BibitemShut {NoStop}%
\bibitem [{\citenamefont {Davies}(1974)}]{Davies1974}%
  \BibitemOpen
  \bibfield  {author} {\bibinfo {author} {\bibfnamefont {E.~B.}\ \bibnamefont
  {Davies}},\ }\href@noop {} {\bibfield  {journal} {\bibinfo  {journal}
  {Commun. Math. Phys.}\ }\textbf {\bibinfo {volume} {39}},\ \bibinfo {pages}
  {91} (\bibinfo {year} {1974})}\BibitemShut {NoStop}%
\bibitem [{\citenamefont {Lindblad}(1976)}]{Lindblad1976}%
  \BibitemOpen
  \bibfield  {author} {\bibinfo {author} {\bibfnamefont {G.}~\bibnamefont
  {Lindblad}},\ }\href@noop {} {\bibfield  {journal} {\bibinfo  {journal}
  {Commun.Math. Phys.}\ }\textbf {\bibinfo {volume} {48}},\ \bibinfo {pages}
  {119} (\bibinfo {year} {1976})}\BibitemShut {NoStop}%
\bibitem [{\citenamefont {Davies}(1976)}]{Davies1976}%
  \BibitemOpen
  \bibfield  {author} {\bibinfo {author} {\bibfnamefont {E.~B.}\ \bibnamefont
  {Davies}},\ }\href@noop {} {\emph {\bibinfo {title} {Quantum {{Theory}} of
  {{Open Systems}}}}}\ (\bibinfo  {publisher} {{Academic Press}},\ \bibinfo
  {year} {1976})\BibitemShut {NoStop}%
\bibitem [{\citenamefont {Wolf}\ and\ \citenamefont {Cirac}(2008)}]{Wolf2008}%
  \BibitemOpen
  \bibfield  {author} {\bibinfo {author} {\bibfnamefont {M.~M.}\ \bibnamefont
  {Wolf}}\ and\ \bibinfo {author} {\bibfnamefont {J.~I.}\ \bibnamefont
  {Cirac}},\ }\href@noop {} {\bibfield  {journal} {\bibinfo  {journal} {Commun.
  Math. Phys.}\ }\textbf {\bibinfo {volume} {279}},\ \bibinfo {pages} {147}
  (\bibinfo {year} {2008})}\BibitemShut {NoStop}%
\bibitem [{\citenamefont {Rivas}\ \emph {et~al.}(2010)\citenamefont {Rivas},
  \citenamefont {Huelga},\ and\ \citenamefont {Plenio}}]{Rivas2010}%
  \BibitemOpen
  \bibfield  {author} {\bibinfo {author} {\bibfnamefont {{\'A}.}~\bibnamefont
  {Rivas}}, \bibinfo {author} {\bibfnamefont {S.~F.}\ \bibnamefont {Huelga}}, \
  and\ \bibinfo {author} {\bibfnamefont {M.~B.}\ \bibnamefont {Plenio}},\
  }\href@noop {} {\bibfield  {journal} {\bibinfo  {journal} {Phys. Rev. Lett.}\
  }\textbf {\bibinfo {volume} {105}},\ \bibinfo {pages} {050403} (\bibinfo
  {year} {2010})}\BibitemShut {NoStop}%
\bibitem [{\citenamefont {Rivas}\ and\ \citenamefont
  {Huelga}(2011)}]{Rivas2011}%
  \BibitemOpen
  \bibfield  {author} {\bibinfo {author} {\bibfnamefont {{\'A}.}~\bibnamefont
  {Rivas}}\ and\ \bibinfo {author} {\bibfnamefont {S.~F.}\ \bibnamefont
  {Huelga}},\ }\href@noop {} {\emph {\bibinfo {title} {Open {{Quantum
  Systems}}: {{An Introduction}}}}}\ (\bibinfo  {publisher} {{Springer-Verlag
  Berlin Heidelberg}},\ \bibinfo {year} {2011})\BibitemShut {NoStop}%
\bibitem [{\citenamefont {Breuer}\ \emph {et~al.}(2009)\citenamefont {Breuer},
  \citenamefont {Laine},\ and\ \citenamefont {Piilo}}]{Breuer2009}%
  \BibitemOpen
  \bibfield  {author} {\bibinfo {author} {\bibfnamefont {H.-P.}\ \bibnamefont
  {Breuer}}, \bibinfo {author} {\bibfnamefont {E.-M.}\ \bibnamefont {Laine}}, \
  and\ \bibinfo {author} {\bibfnamefont {J.}~\bibnamefont {Piilo}},\
  }\href@noop {} {\bibfield  {journal} {\bibinfo  {journal} {Phys. Rev. Lett.}\
  }\textbf {\bibinfo {volume} {103}},\ \bibinfo {pages} {210401} (\bibinfo
  {year} {2009})}\BibitemShut {NoStop}%
\bibitem [{\citenamefont {Breuer}\ \emph {et~al.}(2016)\citenamefont {Breuer},
  \citenamefont {Laine}, \citenamefont {Piilo},\ and\ \citenamefont
  {Vacchini}}]{Breuer2016}%
  \BibitemOpen
  \bibfield  {author} {\bibinfo {author} {\bibfnamefont {H.-P.}\ \bibnamefont
  {Breuer}}, \bibinfo {author} {\bibfnamefont {E.-M.}\ \bibnamefont {Laine}},
  \bibinfo {author} {\bibfnamefont {J.}~\bibnamefont {Piilo}}, \ and\ \bibinfo
  {author} {\bibfnamefont {B.}~\bibnamefont {Vacchini}},\ }\href@noop {}
  {\bibfield  {journal} {\bibinfo  {journal} {Rev. Mod. Phys.}\ }\textbf
  {\bibinfo {volume} {88}},\ \bibinfo {pages} {021002} (\bibinfo {year}
  {2016})}\BibitemShut {NoStop}%
\bibitem [{\citenamefont {Bylicka}\ \emph {et~al.}(2017)\citenamefont
  {Bylicka}, \citenamefont {Johansson},\ and\ \citenamefont
  {Ac{\'\i}n}}]{Bylicka2017}%
  \BibitemOpen
  \bibfield  {author} {\bibinfo {author} {\bibfnamefont {B.}~\bibnamefont
  {Bylicka}}, \bibinfo {author} {\bibfnamefont {M.}~\bibnamefont {Johansson}},
  \ and\ \bibinfo {author} {\bibfnamefont {A.}~\bibnamefont {Ac{\'\i}n}},\
  }\href@noop {} {\bibfield  {journal} {\bibinfo  {journal} {Phys. Rev. Lett.}\
  }\textbf {\bibinfo {volume} {118}},\ \bibinfo {pages} {120501} (\bibinfo
  {year} {2017})}\BibitemShut {NoStop}%
\bibitem [{\citenamefont {Buscemi}\ and\ \citenamefont
  {Datta}(2016)}]{Buscemi2016}%
  \BibitemOpen
  \bibfield  {author} {\bibinfo {author} {\bibfnamefont {F.}~\bibnamefont
  {Buscemi}}\ and\ \bibinfo {author} {\bibfnamefont {N.}~\bibnamefont
  {Datta}},\ }\href@noop {} {\bibfield  {journal} {\bibinfo  {journal} {Phys.
  Rev. A}\ }\textbf {\bibinfo {volume} {93}},\ \bibinfo {pages} {012101}
  (\bibinfo {year} {2016})}\BibitemShut {NoStop}%
\bibitem [{\citenamefont {Chru{\'s}ci{\'n}ski}\ and\ \citenamefont
  {Rivas}(2017)}]{Chruscinski2017a}%
  \BibitemOpen
  \bibfield  {author} {\bibinfo {author} {\bibfnamefont {D.}~\bibnamefont
  {Chru{\'s}ci{\'n}ski}}\ and\ \bibinfo {author} {\bibfnamefont
  {{\'A}.}~\bibnamefont {Rivas}},\ }\href@noop {} {\bibfield  {journal}
  {\bibinfo  {journal} {arXiv:1710.06771 [quant-ph]}\ } (\bibinfo {year}
  {2017})}\BibitemShut {NoStop}%
\bibitem [{\citenamefont {Lu}\ \emph {et~al.}(2010)\citenamefont {Lu},
  \citenamefont {Wang},\ and\ \citenamefont {Sun}}]{Lu2010}%
  \BibitemOpen
  \bibfield  {author} {\bibinfo {author} {\bibfnamefont {X.-M.}\ \bibnamefont
  {Lu}}, \bibinfo {author} {\bibfnamefont {X.}~\bibnamefont {Wang}}, \ and\
  \bibinfo {author} {\bibfnamefont {C.~P.}\ \bibnamefont {Sun}},\ }\href@noop
  {} {\bibfield  {journal} {\bibinfo  {journal} {Phys. Rev. A}\ }\textbf
  {\bibinfo {volume} {82}},\ \bibinfo {pages} {042103} (\bibinfo {year}
  {2010})}\BibitemShut {NoStop}%
\bibitem [{\citenamefont {Rajagopal}\ \emph {et~al.}(2010)\citenamefont
  {Rajagopal}, \citenamefont {Usha~Devi},\ and\ \citenamefont
  {Rendell}}]{Rajagopal2010}%
  \BibitemOpen
  \bibfield  {author} {\bibinfo {author} {\bibfnamefont {A.~K.}\ \bibnamefont
  {Rajagopal}}, \bibinfo {author} {\bibfnamefont {A.~R.}\ \bibnamefont
  {Usha~Devi}}, \ and\ \bibinfo {author} {\bibfnamefont {R.~W.}\ \bibnamefont
  {Rendell}},\ }\href@noop {} {\bibfield  {journal} {\bibinfo  {journal} {Phys.
  Rev. A}\ }\textbf {\bibinfo {volume} {82}},\ \bibinfo {pages} {042107}
  (\bibinfo {year} {2010})}\BibitemShut {NoStop}%
\bibitem [{\citenamefont {Luo}\ \emph {et~al.}(2012)\citenamefont {Luo},
  \citenamefont {Fu},\ and\ \citenamefont {Song}}]{Luo2012}%
  \BibitemOpen
  \bibfield  {author} {\bibinfo {author} {\bibfnamefont {S.}~\bibnamefont
  {Luo}}, \bibinfo {author} {\bibfnamefont {S.}~\bibnamefont {Fu}}, \ and\
  \bibinfo {author} {\bibfnamefont {H.}~\bibnamefont {Song}},\ }\href@noop {}
  {\bibfield  {journal} {\bibinfo  {journal} {Phys. Rev. A}\ }\textbf {\bibinfo
  {volume} {86}},\ \bibinfo {pages} {044101} (\bibinfo {year}
  {2012})}\BibitemShut {NoStop}%
\bibitem [{\citenamefont {Bylicka}\ \emph {et~al.}(2014)\citenamefont
  {Bylicka}, \citenamefont {Chru{\'s}ci{\'n}ski},\ and\ \citenamefont
  {Maniscalco}}]{Bylicka2014}%
  \BibitemOpen
  \bibfield  {author} {\bibinfo {author} {\bibfnamefont {B.}~\bibnamefont
  {Bylicka}}, \bibinfo {author} {\bibfnamefont {D.}~\bibnamefont
  {Chru{\'s}ci{\'n}ski}}, \ and\ \bibinfo {author} {\bibfnamefont
  {S.}~\bibnamefont {Maniscalco}},\ }\href@noop {} {\bibfield  {journal}
  {\bibinfo  {journal} {Sci. Rep.}\ }\textbf {\bibinfo {volume} {4}},\ \bibinfo
  {pages} {5720} (\bibinfo {year} {2014})}\BibitemShut {NoStop}%
\bibitem [{\citenamefont {Lorenzo}\ \emph {et~al.}(2013)\citenamefont
  {Lorenzo}, \citenamefont {Plastina},\ and\ \citenamefont
  {Paternostro}}]{Lorenzo2013}%
  \BibitemOpen
  \bibfield  {author} {\bibinfo {author} {\bibfnamefont {S.}~\bibnamefont
  {Lorenzo}}, \bibinfo {author} {\bibfnamefont {F.}~\bibnamefont {Plastina}}, \
  and\ \bibinfo {author} {\bibfnamefont {M.}~\bibnamefont {Paternostro}},\
  }\href@noop {} {\bibfield  {journal} {\bibinfo  {journal} {Phys. Rev. A}\
  }\textbf {\bibinfo {volume} {88}},\ \bibinfo {pages} {020102} (\bibinfo
  {year} {2013})}\BibitemShut {NoStop}%
\bibitem [{\citenamefont {Dhar}\ \emph {et~al.}(2015)\citenamefont {Dhar},
  \citenamefont {Bera},\ and\ \citenamefont {Adesso}}]{Dhar2015}%
  \BibitemOpen
  \bibfield  {author} {\bibinfo {author} {\bibfnamefont {H.~S.}\ \bibnamefont
  {Dhar}}, \bibinfo {author} {\bibfnamefont {M.~N.}\ \bibnamefont {Bera}}, \
  and\ \bibinfo {author} {\bibfnamefont {G.}~\bibnamefont {Adesso}},\
  }\href@noop {} {\bibfield  {journal} {\bibinfo  {journal} {Phys. Rev. A}\
  }\textbf {\bibinfo {volume} {91}},\ \bibinfo {pages} {032115} (\bibinfo
  {year} {2015})}\BibitemShut {NoStop}%
\bibitem [{\citenamefont {Daffer}\ \emph {et~al.}(2004)\citenamefont {Daffer},
  \citenamefont {W{\'o}dkiewicz}, \citenamefont {Cresser},\ and\ \citenamefont
  {McIver}}]{Daffer2004}%
  \BibitemOpen
  \bibfield  {author} {\bibinfo {author} {\bibfnamefont {S.}~\bibnamefont
  {Daffer}}, \bibinfo {author} {\bibfnamefont {K.}~\bibnamefont
  {W{\'o}dkiewicz}}, \bibinfo {author} {\bibfnamefont {J.~D.}\ \bibnamefont
  {Cresser}}, \ and\ \bibinfo {author} {\bibfnamefont {J.~K.}\ \bibnamefont
  {McIver}},\ }\href@noop {} {\bibfield  {journal} {\bibinfo  {journal} {Phys.
  Rev. A}\ }\textbf {\bibinfo {volume} {70}},\ \bibinfo {pages} {010304}
  (\bibinfo {year} {2004})}\BibitemShut {NoStop}%
\bibitem [{\citenamefont {Shabani}\ and\ \citenamefont
  {Lidar}(2005)}]{Shabani2005}%
  \BibitemOpen
  \bibfield  {author} {\bibinfo {author} {\bibfnamefont {A.}~\bibnamefont
  {Shabani}}\ and\ \bibinfo {author} {\bibfnamefont {D.~A.}\ \bibnamefont
  {Lidar}},\ }\href@noop {} {\bibfield  {journal} {\bibinfo  {journal} {Phys.
  Rev. A}\ }\textbf {\bibinfo {volume} {71}},\ \bibinfo {pages} {020101}
  (\bibinfo {year} {2005})}\BibitemShut {NoStop}%
\bibitem [{\citenamefont {Rivas}\ \emph {et~al.}(2014)\citenamefont {Rivas},
  \citenamefont {Huelga},\ and\ \citenamefont {Plenio}}]{Rivas2014}%
  \BibitemOpen
  \bibfield  {author} {\bibinfo {author} {\bibfnamefont {{\'A}.}~\bibnamefont
  {Rivas}}, \bibinfo {author} {\bibfnamefont {S.~F.}\ \bibnamefont {Huelga}}, \
  and\ \bibinfo {author} {\bibfnamefont {M.~B.}\ \bibnamefont {Plenio}},\
  }\href@noop {} {\bibfield  {journal} {\bibinfo  {journal} {Rep. Prog. Phys.}\
  }\textbf {\bibinfo {volume} {77}},\ \bibinfo {pages} {094001} (\bibinfo
  {year} {2014})}\BibitemShut {NoStop}%
\bibitem [{\citenamefont {Koch}(2016)}]{Koch2016}%
  \BibitemOpen
  \bibfield  {author} {\bibinfo {author} {\bibfnamefont {C.~P.}\ \bibnamefont
  {Koch}},\ }\href@noop {} {\bibfield  {journal} {\bibinfo  {journal} {J.
  Phys.: Condens. Matter}\ }\textbf {\bibinfo {volume} {28}},\ \bibinfo {pages}
  {213001} (\bibinfo {year} {2016})}\BibitemShut {NoStop}%
\bibitem [{\citenamefont {{de Vega}}\ and\ \citenamefont
  {Alonso}(2017)}]{deVega2017}%
  \BibitemOpen
  \bibfield  {author} {\bibinfo {author} {\bibfnamefont {I.}~\bibnamefont {{de
  Vega}}}\ and\ \bibinfo {author} {\bibfnamefont {D.}~\bibnamefont {Alonso}},\
  }\href@noop {} {\bibfield  {journal} {\bibinfo  {journal} {Rev. Mod. Phys.}\
  }\textbf {\bibinfo {volume} {89}},\ \bibinfo {pages} {015001} (\bibinfo
  {year} {2017})}\BibitemShut {NoStop}%
\bibitem [{\citenamefont {Breuer}\ and\ \citenamefont
  {Petruccione}(2007)}]{Breuer2007a}%
  \BibitemOpen
  \bibfield  {author} {\bibinfo {author} {\bibfnamefont {H.-P.}\ \bibnamefont
  {Breuer}}\ and\ \bibinfo {author} {\bibfnamefont {F.}~\bibnamefont
  {Petruccione}},\ }\href@noop {} {\emph {\bibinfo {title} {The {{Theory}} of
  {{Open Quantum Systems}}}}}\ (\bibinfo  {publisher} {{Oxford University
  Press}},\ \bibinfo {address} {Oxford},\ \bibinfo {year} {2007})\BibitemShut
  {NoStop}%
\bibitem [{\citenamefont {Gardiner}\ and\ \citenamefont
  {Zoller}(2004)}]{Gardiner2010}%
  \BibitemOpen
  \bibfield  {author} {\bibinfo {author} {\bibfnamefont {C.}~\bibnamefont
  {Gardiner}}\ and\ \bibinfo {author} {\bibfnamefont {P.}~\bibnamefont
  {Zoller}},\ }\href@noop {} {\emph {\bibinfo {title} {Quantum {{Noise}}: {{A
  Handbook}} of {{Markovian}} and {{Non}}-{{Markovian Quantum Stochastic
  Methods}} with {{Applications}} to {{Quantum Optics}}}}}\ (\bibinfo
  {publisher} {{Springer-Verlag Berlin Heidelberg}},\ \bibinfo {year}
  {2004})\BibitemShut {NoStop}%
\bibitem [{\citenamefont {Holevo}(1973)}]{Holevo1973}%
  \BibitemOpen
  \bibfield  {author} {\bibinfo {author} {\bibfnamefont {A.~S.}\ \bibnamefont
  {Holevo}},\ }\href@noop {} {\bibfield  {journal} {\bibinfo  {journal} {J.
  Multivar. Anal.}\ }\textbf {\bibinfo {volume} {3}},\ \bibinfo {pages} {337}
  (\bibinfo {year} {1973})}\BibitemShut {NoStop}%
\bibitem [{\citenamefont {Helstrom}(1976)}]{Helstrom1976}%
  \BibitemOpen
  \bibfield  {author} {\bibinfo {author} {\bibfnamefont {C.~W.}\ \bibnamefont
  {Helstrom}},\ }\href@noop {} {\emph {\bibinfo {title} {Quantum {{Detection}}
  and {{Estimation Theory}}}}}\ (\bibinfo  {publisher} {{Academic Press}},\
  \bibinfo {address} {New York},\ \bibinfo {year} {1976})\BibitemShut {NoStop}%
\end{thebibliography}%
\newpage \setcounter{equation}{0} \setcounter{section}{0} \setcounter{subsection}{0} \renewcommand{\theequation}{S\arabic{equation}} \onecolumngrid \setcounter{enumiv}{0}

\section*{Appendices}

\section{Evolution of the system}

\subsection{Solution}

In this problem, a system qubit is interacted with a bath qubit via
the following $X-X$ coupling,
\begin{equation}
\hi=\xi\xs\otimes\xb.
\end{equation}
To simulate the restoration of the bath state, we apply a cooling
process on the bath qubit to continuously transit the bath state to
the initial state of the bath $\ket[\ob]$, which can be considered
as an energy relaxation process and can be described by $\kappa\lb\rsb\delt$
\cite{Breuer2007a,Gardiner2010}, where $\kappa$ is the cooling rate,
$\rsb$ is the density matrix of the joint state of the system and
bath at time $t$, and $\lb$ is the dissipator defined as
\begin{equation}
\lb\rsb=\sm\rsb\sp-\frac{1}{2}\{\sp\sm,\,\rsb\}.
\end{equation}
Putting the Hamiltonian evolution and the cooling process together,
we have the following master equation for the joint state of the system
qubit and the bath qubit:
\begin{equation}
\pt\rsb=-\i\xi[\xs\otimes\xb,\rsb]+\kappa\lb\rsb.\label{eq:master eq-1}
\end{equation}

Suppose the initial density matrix of the system qubit is $\rs[0]=\frac{1}{2}\big(\is+\xo\xs+\yo\ys+\zo\zs\big)$,
and the initial state of the bath qubit is $\ket[\ob]$. To solve
the master equation (\ref{eq:master eq-1}), we work in the operator
space of the system and bath qubits, which has $\si\otimes\sj$, $i,j=0,x,y,z$,
as its basis, where $\sigma_{0}=I$. In this representation, the joint
density matrix of the system and the bath can be represented by a
$16\times1$ vector, and a superoperator on the system and the bath
can be represented by a $16\times16$ matrix. The vector form of the
initial density matrix of the system and the bath is
\[
\vsb[0]=\frac{1}{4}[1,0,0,-1,x_{0},0,0,-x_{0},y_{0},0,0,-y_{0},z_{0},0,0,-z_{0}]^{T}.
\]
The master equation (\ref{eq:master eq-1}) can be written in a matrix
form:
\begin{equation}
\pt\vsb=M\vsb,\label{eq:vec mas eq}
\end{equation}
where
\begin{equation}
M=\left(\begin{array}{cccccccccccccccc}
0 & 0 & 0 & 0 & 0 & 0 & 0 & 0 & 0 & 0 & 0 & 0 & 0 & 0 & 0 & 0\\
0 & -\frac{\kappa}{2} & 0 & 0 & 0 & 0 & 0 & 0 & 0 & 0 & 0 & 0 & 0 & 0 & 0 & 0\\
0 & 0 & -\frac{\kappa}{2} & 0 & 0 & 0 & 0 & -2\xi & 0 & 0 & 0 & 0 & 0 & 0 & 0 & 0\\
-\kappa & 0 & 0 & -\kappa & 0 & 0 & 2\xi & 0 & 0 & 0 & 0 & 0 & 0 & 0 & 0 & 0\\
0 & 0 & 0 & 0 & 0 & 0 & 0 & 0 & 0 & 0 & 0 & 0 & 0 & 0 & 0 & 0\\
0 & 0 & 0 & 0 & 0 & -\frac{\kappa}{2} & 0 & 0 & 0 & 0 & 0 & 0 & 0 & 0 & 0 & 0\\
0 & 0 & 0 & -2\xi & 0 & 0 & -\frac{\kappa}{2} & 0 & 0 & 0 & 0 & 0 & 0 & 0 & 0 & 0\\
0 & 0 & 2\xi & 0 & -\kappa & 0 & 0 & -\kappa & 0 & 0 & 0 & 0 & 0 & 0 & 0 & 0\\
0 & 0 & 0 & 0 & 0 & 0 & 0 & 0 & 0 & 0 & 0 & 0 & 0 & -2\xi & 0 & 0\\
0 & 0 & 0 & 0 & 0 & 0 & 0 & 0 & 0 & -\frac{\kappa}{2} & 0 & 0 & -2\xi & 0 & 0 & 0\\
0 & 0 & 0 & 0 & 0 & 0 & 0 & 0 & 0 & 0 & -\frac{\kappa}{2} & 0 & 0 & 0 & 0 & 0\\
0 & 0 & 0 & 0 & 0 & 0 & 0 & 0 & -\kappa & 0 & 0 & -\kappa & 0 & 0 & 0 & 0\\
0 & 0 & 0 & 0 & 0 & 0 & 0 & 0 & 0 & 2\xi & 0 & 0 & 0 & 0 & 0 & 0\\
0 & 0 & 0 & 0 & 0 & 0 & 0 & 0 & 2\xi & 0 & 0 & 0 & 0 & -\frac{\kappa}{2} & 0 & 0\\
0 & 0 & 0 & 0 & 0 & 0 & 0 & 0 & 0 & 0 & 0 & 0 & 0 & 0 & -\frac{\kappa}{2} & 0\\
0 & 0 & 0 & 0 & 0 & 0 & 0 & 0 & 0 & 0 & 0 & 0 & -\kappa & 0 & 0 & -\kappa
\end{array}\right).
\end{equation}
The solution to $\vsb$ can be obtain by solving Eq. (\ref{eq:vec mas eq})
directly:
\begin{equation}
\vsb=\exp(Mt)\vsb[0].
\end{equation}

If we trace out the freedom degrees of the bath qubit, the reduced
density matrix of the system qubit at time $t$ is
\begin{equation}
\rs=\frac{1}{2}\Big[I+x_{0}\xs+\ct\Big(\yo\ys+\zo\zs\Big)\Big],\label{eq:sta evo}
\end{equation}
where
\begin{equation}
\ct=\e^{-\frac{\kappa t}{4}}\bigg(\frac{\kappa\sinh\frac{t}{4}\sqrt{\kappa^{2}-64\xi^{2}}}{\sqrt{\kappa^{2}-64\xi^{2}}}+\cosh\frac{t}{4}\sqrt{\kappa^{2}-64\xi^{2}}\bigg).\label{eq:ct}
\end{equation}

Eq. (\ref{eq:ct}) implicitly assumes that $\kappa^{2}>64\xi^{2}$.
If $\kappa^{2}<64\xi^{2}$, $\ct$ becomes
\begin{equation}
\ct=\e^{-\frac{\kappa t}{4}}\bigg(\frac{\kappa\sin\frac{t}{4}\sqrt{64\xi^{2}-\kappa^{2}}}{\sqrt{64\xi^{2}-\kappa^{2}}}+\cos\frac{t}{4}\sqrt{64\xi^{2}-\kappa^{2}}\bigg).\label{eq:ct2}
\end{equation}
If $\kappa^{2}=64\xi^{2}$, $\ct$ is
\begin{equation}
\ct=\e^{-\frac{\kappa t}{4}}\Big(1+\frac{1}{4}\kappa t\Big).\label{eq:ct3}
\end{equation}

\subsection{Master equation}

If we want to determine the Markovianity of the system dynamics based
on the CP-divisibility criterion, we need to obtain the master equation
for the system qubit and check whether the coefficients of the dissipation
terms are non-negative or not \cite{Rivas2011}. Generally, deriving
the master equation for an open quantum system is not easy, but in
this problem, as we have obtained the solution to the system evolution,
we can infer the master equation of the system from that solution.

To derive the master equation of the system, we transform into the
operator space of the system which has basis $\{I,\xs,\ys,\zs\}$.
The density matrix of the system can be represented by a $4\times1$
vector, and a superoperator on the system can be represented by a
$4\times4$ matrix. In this representation, the vector form of the
system qubit at time $t$ is
\begin{equation}
\vs=\frac{1}{2}[\wo,\xo,\ct\yo,\ct\zo]^{T}=\qt\vs[0],
\end{equation}
where $\vs[0]=\frac{1}{2}[\wo,\xo,\yo,\zo]^{T}$ is the vector form
of the initial density matrix of the system qubit, and $\qt$ is the
matrix representation of the system qubit evolution from the initial
time to the time $t$,
\begin{equation}
\qt=\left(\begin{array}{cccc}
1 & 0 & 0 & 0\\
0 & 1 & 0 & 0\\
0 & 0 & c_{t} & 0\\
0 & 0 & 0 & c_{t}
\end{array}\right).
\end{equation}
Note that $\wo=1$ for a normalized density matrix, but as we want
to find the linear transformation corresponding to the time change
of $\vs$ in the operator space of the system, we temporarily denote
it as a variable. It will be restored to $1$ when the linear transformation
is derived. 

The time change of $\vs$ is
\begin{equation}
\pt\vs=\dqt\vs[0]=\dqt\qt[][-1]\vs.
\end{equation}
Thus, $\dqt\qt[][-1]$ is the matrix representation of the linear
transformation corresponding to the time derivative of the system
density matrix, and
\begin{equation}
\dqt\qt[][-1]=\left(\begin{array}{cccc}
0 & 0 & 0 & 0\\
0 & 0 & 0 & 0\\
0 & 0 & \frac{\dct}{\ct} & 0\\
0 & 0 & 0 & \frac{\dct}{\ct}
\end{array}\right).\label{eq:qtqinvers}
\end{equation}

Now we can find the superoperator corresponding to $\dqt\qt[][-1]$.
In order to do this, we need to know the matrix representations $\sij$
for the basis of the superoperator $\si{}[\cdot]\sj[\s]$. It is straightforward
to obtain that
\begin{equation}
\begin{array}{cccc}
\text{s}_{00}=\left(\begin{array}{cccc}
1 & 0 & 0 & 0\\
0 & 1 & 0 & 0\\
0 & 0 & 1 & 0\\
0 & 0 & 0 & 1
\end{array}\right), & \text{s}_{0x}=\left(\begin{array}{cccc}
0 & 1 & 0 & 0\\
1 & 0 & 0 & 0\\
0 & 0 & 0 & i\\
0 & 0 & -i & 0
\end{array}\right), & \text{s}_{0y}=\left(\begin{array}{cccc}
0 & 0 & 1 & 0\\
0 & 0 & 0 & -i\\
1 & 0 & 0 & 0\\
0 & i & 0 & 0
\end{array}\right), & \text{s}_{0z}=\left(\begin{array}{cccc}
0 & 0 & 0 & 1\\
0 & 0 & i & 0\\
0 & -i & 0 & 0\\
1 & 0 & 0 & 0
\end{array}\right),\\
\text{s}_{x0}=\left(\begin{array}{cccc}
0 & 1 & 0 & 0\\
1 & 0 & 0 & 0\\
0 & 0 & 0 & -i\\
0 & 0 & i & 0
\end{array}\right), & \text{s}_{xx}=\left(\begin{array}{cccc}
1 & 0 & 0 & 0\\
0 & 1 & 0 & 0\\
0 & 0 & -1 & 0\\
0 & 0 & 0 & -1
\end{array}\right), & \text{s}_{xy}=\left(\begin{array}{cccc}
0 & 0 & 0 & -i\\
0 & 0 & 1 & 0\\
0 & 1 & 0 & 0\\
i & 0 & 0 & 0
\end{array}\right), & \text{s}_{xz}=\left(\begin{array}{cccc}
0 & 0 & i & 0\\
0 & 0 & 0 & 1\\
-i & 0 & 0 & 0\\
0 & 1 & 0 & 0
\end{array}\right),\\
\text{s}_{y0}=\left(\begin{array}{cccc}
0 & 0 & 1 & 0\\
0 & 0 & 0 & i\\
1 & 0 & 0 & 0\\
0 & -i & 0 & 0
\end{array}\right), & \text{s}_{yx}=\left(\begin{array}{cccc}
0 & 0 & 0 & i\\
0 & 0 & 1 & 0\\
0 & 1 & 0 & 0\\
-i & 0 & 0 & 0
\end{array}\right), & \text{s}_{yy}=\left(\begin{array}{cccc}
1 & 0 & 0 & 0\\
0 & -1 & 0 & 0\\
0 & 0 & 1 & 0\\
0 & 0 & 0 & -1
\end{array}\right), & \text{s}_{yz}=\left(\begin{array}{cccc}
0 & -i & 0 & 0\\
i & 0 & 0 & 0\\
0 & 0 & 0 & 1\\
0 & 0 & 1 & 0
\end{array}\right),\\
\text{s}_{z0}=\left(\begin{array}{cccc}
0 & 0 & 0 & 1\\
0 & 0 & -i & 0\\
0 & i & 0 & 0\\
1 & 0 & 0 & 0
\end{array}\right), & \text{s}_{zx}=\left(\begin{array}{cccc}
0 & 0 & -i & 0\\
0 & 0 & 0 & 1\\
i & 0 & 0 & 0\\
0 & 1 & 0 & 0
\end{array}\right), & \text{s}_{zy}=\left(\begin{array}{cccc}
0 & i & 0 & 0\\
-i & 0 & 0 & 0\\
0 & 0 & 0 & 1\\
0 & 0 & 1 & 0
\end{array}\right), & \text{s}_{zz}=\left(\begin{array}{cccc}
1 & 0 & 0 & 0\\
0 & -1 & 0 & 0\\
0 & 0 & -1 & 0\\
0 & 0 & 0 & 1
\end{array}\right).
\end{array}
\end{equation}
By decomposing $\dqt\qt[][-1]$ in Eq. (\ref{eq:qtqinvers}) along
$\sij$, one can obtain that
\begin{equation}
\dqt\qt[][-1]=\frac{\dct}{2\ct}(\sij[00]-\sij[xx]).
\end{equation}
Therefore, we have
\begin{equation}
\pt\rs=\frac{\dct}{2\ct}(\rs-\xs\rs\xs)=-\frac{\dct}{2\ct}\lb[\xs]\rs,
\end{equation}
which is the master equation for the system qubit.

\section{Non-Markovianity measure based on trace distance}

A useful measure of non-Markovianity of a quantum dynamics was proposed
by Breuer \emph{et al.} \cite{Breuer2009} from an information theoretic
point of view. The measure is the integral of the increase in the
trace distance between two quantum states under that quantum dynamics
maximized over all possible pairs of initial quantum states.

Mathematically, if we denote the measure of non-Markovianity as $\mn$,
and two quantum states under the quantum dynamics of interest as $\rk[1]$
and $\rk[2]$, the measure of non-Markovianity is
\begin{equation}
\mn=\max_{\rk[1,2][0]}\int_{\pt[\tp]d_{\tr}>0}\pt[\tp]\dtr[{\rk[1][\tp]}][{\rk[2][\tp]}]\dtp,
\end{equation}
where $\dtr[{\rk[1][\tp]}][{\rk[2][\tp]}]$ is the trace distance
between $\rk[1][\tp]$ and $\rk[2][\tp]$ defined as
\begin{equation}
\dtr[{\rk[1][\tp]}][{\rk[2][\tp]}]=\frac{1}{2}\trd{{{\rk[1][\tp]}}-{{\rk[2][\tp]}}},
\end{equation}
and the integral is over all time intervals where the trace distance
between $\rk[1][\tp]$, $\rk[2][\tp]$ increases, i.e., $\pt[\tp]\dtr[{\rk[1][\tp]}][{\rk[2][\tp]}]>0$. 

In this problem, we assume we have two initial states for the system
qubit,
\begin{equation}
\begin{aligned}\rs[0]= & \frac{1}{2}(\is+\xo\xs+\yo\ys+\zo\zs),\\
\rsp[0]= & \frac{1}{2}(\is+\xop\xs+\yop\ys+\zop\zs).
\end{aligned}
\end{equation}
Then, according to Eq. (\ref{eq:sta evo}), the difference between
$\rs$ and $\rsp$ at time $t$ is
\begin{equation}
\rs-\rsp=\frac{1}{2}\Big(\dx\xs+\ct\dy\ys+\ct\dz\zs\Big),
\end{equation}
and the trace distance between $\rs$ and $\rsp$ is
\begin{equation}
\dtr[{\rs}][{\rsp}]=\frac{1}{2}\sqrt{\dx[2]+\cts\left(\dy[2]+\dz[2]\right)},\label{eq:tr dist}
\end{equation}
where $\dx=\xo-\xop$, $\dy=\yo-\yop$, $\dz=\zo-\zop$.

According to the result in the main paper, the trace distance between
two quantum states under the dynamics in this problem may oscillate
only when $\kappa^{2}\leq64\xi^{2}$, so we will focus on this case
below.

To investigate when the trace distance between $\rs$ and $\rsp$
increases, we take the derivative of Eq. (\ref{eq:tr dist}) with
respect to time $t$, which produces
\begin{equation}
\pt\dtr[{\rs}][{\rsp}]=\left(\dy[2]+\dz[2]\right)\frac{\cts}{\dtr[{\rs}][{\rsp}]}\frac{\dct}{\ct}.
\end{equation}
And when $\kappa^{2}<64\xi^{2}$,
\begin{equation}
\frac{\dct}{\ct}=\frac{-16\xi^{2}}{\kappa+\sqrt{64\xi^{2}-\kappa^{2}}\cot\left(\frac{1}{4}t\sqrt{64\xi^{2}-\kappa^{2}}\right)}.
\end{equation}
Obviously, $\frac{\dct}{\ct}>0$ when $\cot\left(\frac{1}{4}t\sqrt{64\xi^{2}-\kappa^{2}}\right)<\frac{-\kappa}{\sqrt{64\xi^{2}-\kappa^{2}}}$,
therefore, the trace distance $\dtr[{\rs}][{\rsp}]$ increases for
any
\begin{equation}
\tn-\delta<t<\tn,\;n=1,2,\cdots,\label{eq:time interv}
\end{equation}
where
\begin{equation}
\tn=\frac{4n\pi}{\sqrt{64\xi^{2}-\kappa^{2}}},\;\delta=\frac{4\arctan\frac{\sqrt{64\xi^{2}-\kappa^{2}}}{\kappa}}{\sqrt{64\xi^{2}-\kappa^{2}}}.
\end{equation}
It can be verified from Eq. (\ref{eq:ct2}) that
\begin{equation}
\ct[{{\tn}}-\delta]=0,\;\ct[{\tn}]=(-1)^{n}\e^{-\frac{\kappa\tn}{4}}.
\end{equation}
Then it follows Eq. (\ref{eq:tr dist}), the increase in the trace
distance in the time interval determined by Eq. (\ref{eq:time interv})
for each $n$ is
\begin{equation}
\dn\dtr[{\rs}][{\rsp}]=\frac{1}{2}\left(\sqrt{\dx[2]+\e^{-\frac{\kappa\tn}{2}}\left(\dy[2]+\dz[2]\right)}-|\dx|\right).
\end{equation}

Now we need to maximize $\dn\dtr[{\rs}][{\rsp}]$ over all possible
$\dx$, $\dy$, $\dz$. Note that $\dn\dtr[{\rs}][{\rsp}]$ can be
written as
\begin{equation}
\dn\dtr[{\rs}][{\rsp}]=\frac{\e^{-\frac{\kappa\tn}{2}}\left(\dy[2]+\dz[2]\right)}{2\left(\sqrt{\dx[2]+\e^{-\frac{\kappa\tn}{2}}\left(\dy[2]+\dz[2]\right)}+|\dx|\right)}.
\end{equation}
If we fix $\dy$ and $\dz$, it is obvious that $\dn\dtr[{\rs}][{\rsp}]$
is maximized when $\dx=0$. And letting $\dx=0$ is possible, because
we can always choose proper $\xo=\xop$ (at least we can choose $\xo=\xop=0$)
without violating the positivity of either $\rs[0]$ or $\rsp[0]$
for any $\dy$ and $\dz$. Therefore,
\begin{equation}
\max_{\dx}\dn\dtr[{\rs}][{\rsp}]=\frac{1}{2}\e^{-\frac{\kappa\tn}{4}}\sqrt{\dy[2]+\dz[2]}.\label{eq:6}
\end{equation}

To maximize Eq. (\ref{eq:6}) over $\dy$ and $\dz$, we denote two
vectors $\vr=[\xo,\yo,\zo]$, $\vrp=[\xo,\yop,\zop]$, and $\|\vr\|\leq1$,
$\|\vrp\|\leq1$ due to the positivity of $\rs[0]$ and $\rsp[0]$.
Note the $x$ components of $\vr$ and $\vrp$ have been explicitly
chosen to be the same. Then,
\begin{equation}
\max_{\dx}\dn\dtr[{\rs}][{\rsp}]=\frac{1}{2}\e^{-\frac{\kappa\tn}{4}}\|\vr-\vrp\|.
\end{equation}
And the maximization of $\max_{\dx}\dn\dtr[{\rs}][{\rsp}]$ over $\dy$
and $\dz$ becomes a maximization over $\vr$ and $\vrp$. Obviously
$\|\vr-\vrp\|$ is maximized when $\vr$ and $\vrp$ are antiparallel
and $\|\vr\|=\|\vrp\|=1$, which means $\vr=-\vrp$ and thus $\xo=-\xo$,
implying $\xo=0$, and the maximum of $\|\vr-\vrp\|$ is $2$. So,
the maximum of $\dn\dtr[{\rs}][{\rsp}]$ is
\begin{equation}
\max\dn\dtr[{\rs}][{\rsp}]=\e^{-\frac{\kappa\tn}{4}}.
\end{equation}

Therefore, the measure of non-Markovianity $\mn$ for $\kappa^{2}\leq64\xi^{2}$
is
\begin{equation}
\mn=\sum_{n=1}^{+\infty}\e^{-\frac{\kappa\tn}{4}}=\sum_{n=1}^{+\infty}\e^{-\frac{n\kappa\pi}{\sqrt{64\xi^{2}-\kappa^{2}}}}=\frac{1}{\exp\Big(\frac{\kappa\pi}{\sqrt{64\xi^{2}-\kappa^{2}}}\Big)-1}.
\end{equation}
It can be seen that when $\kappa=0$, $\mn$ is infinity, indicating
the maximum non-Markovianity, as the state of the system qubit always
oscillates in this case which induces periodic information backflow
from the bath to the system and the amplitude of the information backflow
never decays. When $\kappa$ decreases, $\mn$ decreases and becomes
finite, indicating a weaker non-Markovianity, as the oscillation of
the system state is weakened and the amplitude of the information
backflow decays with time. When $\kappa^{2}=64\xi^{2}$, $\mn=0$,
indicating no non-Markovianity in the system dynamics.

Of course, if $\kappa^{2}>64\xi^{2}$, as there is no increase in
the trace distance between any two quantum states at any time in this
problem, $\mn$ is zero by its definition for this case. 

\section{Bath correlation}

In this section, we use the standard procedures under the Born-Markov
approximation \cite{Breuer2007a} to compute the time correlation
functions for the bath qubit. The purpose is to find how fast the
bath correlation must decay in order to validate the Born-Markov approximation
and compare it to the threshold for the transition between Markovian
and non-Markovian dynamics found in our problem.

We first make the following transformation to eliminate the dissipation
term in the master equation (\ref{eq:master eq-1}),
\begin{equation}
\rsbt=\e^{-\kappa t\lb}\rsb.\label{eq:transform}
\end{equation}
Then,
\begin{equation}
\pt\rsbt=-\i\xi\e^{-\kappa t\lb}[\xs\otimes\xb,\rsb].\label{eq:0}
\end{equation}
In order to make it an equation for $\rsbt$, we write $\rsb$ as
$\rsb=\e^{\kappa t\lb}\rsbt$, then
\begin{equation}
\pt\rsbt=-\i\xi\e^{-\kappa t\lb}[\xs\otimes\xb,\e^{\kappa t\lb}\rsbt].\label{eq:1}
\end{equation}
If we integrate Eq. (\ref{eq:1}) over the time, we have
\begin{equation}
\rsbt=\rsbt[0]-\i\xi\int_{0}^{t}\e^{-\kappa\tp\lb}[\xs\otimes\xb,\e^{\kappa\tp\lb}\rsbt[\tp]]\dtp.\label{eq:2}
\end{equation}
Plugging Eq. (\ref{eq:2}) into Eq. (\ref{eq:0}), we can obtain a
master equation of $\rsbt$ up to the second order of $\xi$,
\begin{equation}
\pt\rsbt=-\i\xi\e^{-\kappa t\lb}[\xs\otimes\xb,\e^{\kappa t\lb}\rsbt[0]]-\xi^{2}\e^{-\kappa t\lb}\int_{0}^{t}[\xs\otimes\xb,\e^{\kappa(t-\tp)\lb}[\xs\otimes\xb,\e^{\kappa\tp\lb}\rsbt[\tp]]]\dtp.\label{eq:3}
\end{equation}
In order to obtain the reduced master equation for the system qubit
alone, we trace out the freedom degrees of the bath qubit for Eq.
(\ref{eq:3}). By invoking the Born approximation, we assume that
\begin{equation}
\rsb\approx\rs\otimes\rb[0],\label{eq:born markov approx}
\end{equation}
where $\rb[0]=\kb[\ob]$. Then, it can be verified that
\begin{equation}
\trb\e^{-\kappa t\lb}[\xs\otimes\xb,\e^{\kappa t\lb}\rsbt[0]]=0.
\end{equation}
So,
\begin{equation}
\pt\rs=\trb\pt\rsbt=-\xi^{2}\int_{0}^{t}\trb[\xs\otimes\xb,\e^{\kappa(t-\tp)\lb}[\xs\otimes\xb,\e^{\kappa\tp\lb}\rsbt[\tp]]]\dtp,\label{eq:4}
\end{equation}
where we have used $\trb\e^{-\kappa t\lb}A=\trb A$ for an arbitrary
operator $A$ of the bath qubit, as $\trb\lb[k]A=0$ for $k\geq1$.
According to the transformation (\ref{eq:transform}) and the Born
approximation (\ref{eq:born markov approx}), Eq. (\ref{eq:4}) can
be simplified to
\begin{equation}
\pt\rs=-\xi^{2}\int_{0}^{t}\trb[\xs\otimes\xb,\e^{\kappa(t-\tp)\lb}[\xs\otimes\xb,\rs[\tp]\otimes\rb[0]]]\dtp,
\end{equation}
followed by
\begin{equation}
\pt\rs=\xi^{2}\int_{0}^{t}\ft\lb[\xs]\rs[\tp]\dtp,\label{eq: 2nd order mas eq}
\end{equation}
where $\ft$ is the bath correlation function
\begin{equation}
\ft=\trb[\xb\e^{\kappa(t-\tp)\lb}(\xb\rb[0]+\rb[0]\xb)].\label{eq:bath corr-1}
\end{equation}
As the initial state of the bath qubit is $\ket[\ob]$, $\rb[0]=\kb[\ob]$,
so $\xb\rb[0]+\rb[0]\xb=\xb$. And it can be easily verified that
$\e^{\kappa(t-\tp)\lb}\xb=\e^{-\frac{\kappa}{2}(t-\tp)}\xb$. Therefore,
\begin{equation}
\ft=\exp\Big[-\frac{1}{2}\kappa(t-\tp)\Big].
\end{equation}
This means that the time width of the bath correlation is $2/\kappa$.

The Markov approximation is to further assume that the bath correlation
$\ft$ decays very fast in the time scale of the system evolution,
so that the $\rs[\tp]$ in the integrand in Eq. (\ref{eq: 2nd order mas eq})
can be replaced by $\rs$ and the evolution of $\rs$ does not depend
on the history of $\rs$ then, which leads a Markovian dynamics of
the system. This essentially requires $\xi\ll\kappa$, in sharp contrast
to the result in the main text that the system dynamics becomes Markovian
when $\kappa\geq8|\xi|$ where $\kappa$ is at the same order of magnitude
of $\xi$.
\end{document}